# Secure management of logs in internet of things

Udit Gupta
Information Networking Institute
Carnegie Mellon University, Pittsburgh-Pennsylvania, USA
uditg@andrew.cmu.edu

**Abstract**

Ever since the advent of computing, managing data has been of extreme importance. With innumerable devices getting added to network infrastructure, there has been a proportionate increase in the data which needs to be stored. With the advent of Internet of Things (IOT) it is anticipated that billions of devices will be a part of the internet in another decade. Since those devices will be communicating with each other on a regular basis with little or no human intervention, plethora of real time data will be generated in quick time which will result in large number of log files. Apart from complexity pertaining to storage, it will be mandatory to maintain confidentiality and integrity of these logs in IOT enabled devices. This paper will provide a brief overview about how logs can be efficiently and securely stored in IOT devices.

## 1. Introduction

Data storage [1 and 2] has been an uphill task for many organizations across the globe. As technology evolved with time, more storage space had to be created which resulted in greater costs incurred by those organizations. As hardware technology evolved from semi-conductor to magnetic to optical storage, the computing power of CPU had to keep pace so as enabling process large data in given time frame. But unfortunately due to limited number of registers, memory hierarchy had to be created like: L1 cache, L2 cache, primary memory, secondary memory. Logs are generally stored in secondary memory and have gained importance in past few years. Logs generally help in debugging, performance review of the system, intrusion detections [4, 5, 6 and 7], creating and analyzing trends based on historical analysis [3]. Up until now, the real time communication between devices was restricted mainly to client-server model and a significant human intervention was involved to generate logs and storing them. But as Internet of Things (IOT) begins to leave an indelible mark in the world of technology, it becomes mandatory to address the issue of generating and storing logs (which will be generated in real time) in an optimized manner so that it becomes easier to access them when required.

IOT [8, 9 and 10] refers to a world of networking where most of the devices will be connected to the internet and will be able to communicate amongst themselves with minimal human intervention. Since the wide range of devices which will be part of IOT infrastructure, their communication will tend to become more complex than traditional networking systems. In order to archive that communication, sufficient memory space needs to be maintained at all times. Given that memory space will be limited especially for devices with limited functionality due to the cost factor, it becomes paramount to categorize logs and then discard all the irrelevant logs.

There are many advantages which logging offers. The biggest advantage being that a system can be debugged and monitored based on the logs stored. Given that as part of organizations, encountering escalations is a common phenomenon which needs to be resolved on a priority basis. Thus, it is important that logs provide sufficient information to developers and researchers so as to ease the process of troubleshooting. Many logging features have been developed keeping in view all the requirements as well as the pace at which technology up gradation is taking place. Five of the most popular logging tools available are: Splunk, Loggly, AlertLogic Log Manager, WhatsUpGold and TIBCO. All these tools offer more or less same functionality like: data indexing [11], real time monitoring, identify trends and patterns for customers, etc. In case of IOT, it is essential that any logging tool deployed must offer to store maximum number of logs in the allotted space since data generated will be huge and the disparity between storage space and available and required will be huge. In this paper, we'll look at some of the efficient methods which can be used for log management in IOT.

## 2. Categorizing logs in IOT enabled devices

Given that there are provisions for almost all entities to get connected to internet, the amount of logs generated will quadruple as compared to cloud networks [12, 13, 14 and 15]. Hence it is important to categorize logs based on their importance in dealing with several performance and network issues. One solution is to divide a part of memory into 5 or 6 segments based on the kind of logs which are encountered by devices and redirect those logs into the corresponding memory unit created and keep those memory units explicitly for those logs. There are six main categories into which logs can be divided: security logs for spam/malware/virus [16, 17 and 18], authentication [19 and 20] related logs pertaining to successful/unsuccessful login, general information logs, logs related to configuration of devices, firewall [21 and 22] logs and device management logs. We'll provide about how these logs can be managed individually and then how to secure them using existing algorithms.

### 2.1 Security logs for malware/spam/virus

Spam, virus and malware are some of the most common security threats in today's time which has prompted developers to come with many algorithms to counter them. But despite taking all precautions, security of the system is often compromised due to penetration of spam, virus or malware. For every spam/malware/virus message being sent to the device, corresponding logs will be generated which needs to be stored. These logs will be helpful in determining trends like from which network these messages are coming from, range of sizes of various messages being sent, similarity in the content, etc. Using these trends we can define a parameter called 'reputation' of a device. This parameter can be determined on a scale of 1-10 with 10 being the highest reputation assigned to device which is in communication with the current device. Also since security of any system is of prime importance, it's recommended that largest a significant fraction of total memory is allotted for these logs.

### 2.2 Authentication related logs

Authentication will play a key role in IOTs since it needs to be ensured that only authorized users gain access to it. The logs related to successful/unsuccessful login attempts needs to be stored which will help forensics department to play a key role in case something goes wrong. Given that users will login to devices multiple times a day, maintaining a long history of login attempts would be infeasible. It must also record all the unsuccessful login attempts so that it can be determined who is attempting to login and how often. Hence it's mandatory to allocate memory for storing these logs.

### 2.3 General information logs

General information needs to be maintained which will include stuff like routes taken for smart cars, grocery items stored for smart refrigerator, etc. These logs will provide user with a holistic picture about the proper functioning of IOT devices. Small amount of space will be required for maintaining these logs and it can be formatted on a daily basis.

### 2.4 Configuration related logs

Any IOT device will need to be configured before its usage. While configuring plethora of information will be generated and this will include various details about the device. Although this information may not seem relevant to end users but it is important to maintain these logs so that any errors pertaining to configuration can be resolved quickly. The space which needs to be allocated for these logs will depend on the devices and the complexity of their configuration.

### 2.5 Firewall logs

In case a firewall support is deployed on IOT device, it is important to maintain logs pertaining to firewall activities which will provide insight into the security breach trying to take place. The information collected via these logs will help in designing robust firewall systems and more efficient firewall policies. It would be better if the space allocated to these logs would be equal to that of authentication logs due to the security aspect.

### 2.6 Device management logs

These logs would contain information about communication with other devices. Any smart device during a certain time period will interact with many other devices. In case if errors are encountered as part of communication between 2 devices then these logs will help identify the root cause and will expedite the process of resolving it. As far as spacing is concerned, it will again depend on how many devices are interacting with the current device.

## 3. Routing and securely storing logs on disk

Logs will carry all the sensitive information pertaining to device and hence it becomes paramount to store them securely in the memory. Many algorithms exist for secure storage of data and most programming languages provide libraries with a wide range of available cryptographic algorithms like Advanced Encryption Standard (AES) [23]. AES-256 with all its security features seems to be the best algorithm available to deploy on IOT devices for securely storing logs in memory. The only overhead which it'll carry with itself is the key management which will consume some amount of memory. This encryption [25] algorithm will ensure that only authorized users are allowed to access them when needed. Let us consider the question of routing the logs generated to appropriate category as mentioned above. To achieve this objective we need to run background processes on the device which will ensure that logs are appropriately routed to their destination. Based on the operating systems designed for IOT [24], scripts can be written corresponding to various logs which can run as background processes periodically similar to task scheduler in windows or cronjob in Linux. Thus, corresponding to six categories of logs mentioned above we can have six different scripts running as background processes which will route logs to appropriate directory. This categorization of logs will ensure that any search on these logs will become more optimized and would give faster results.

In order to save logs onto their own explicit memory, it is necessary to divide total memory allocated for logs into 6 units corresponding to each category of logs as mentioned above. Let us name those memory units as: $mem\_log1$, $mem\_log2$, $mem\_log3$, $mem\_log4$, $mem\_log5$ and $mem\_log6$ and total memory allocated for logs as M' where $mem\_log*$ memory series is subset of M'. The size of each of these memory units will vary depending on the device and how vulnerable it is towards attacks. Each of these memory units can be assigned a threshold value which will serve as an indicator to the user about the space available. Based on these values at any particular time, old logs can be deleted or archived to some other machine based on the memory available. Furthermore, the memory units assigned to each category of logs would not be static and if required can be expanded. Table 1 provides a brief overview about a possible scenario (for 3 smart devices) where row indicates smart devices and column indicates memory corresponding to various logs. Red color indicates if memory consumption is 'above threshold', green indicates 'below threshold' and yellow indicates 'at threshold'. These threshold values can be observed via any monitoring [26, 27, 28 and 29] system.

Table 1

|  | mem_log1 | mem_log2 | mem_log3 | mem_log4 | mem_log5 | mem_log6 |
|---|---|---|---|---|---|---|
| Smart device 1 | Above threshold | At threshold | Below threshold | At threshold | Below threshold | At threshold |
| Smart device 2 | Below threshold | Below threshold | Below threshold | At threshold | Below threshold | At threshold |
| Smart device 3 | Below threshold | Above threshold | Below threshold | Above threshold | At threshold | Below threshold |

4. **Conclusion**

As internet of things continues to pervade through everyday sphere of our life, it also brings along with it the overhead of enormous streams of data generated over a period of time. Unlike cloud infrastructure, the cost of maintaining a datacenter to manage various modules would be extremely high. Thus it is important to manage data locally on the device so that the process of searching over that data would expedite. Although the process of building logging tools for IOT is yet to initiate but given the huge volume of data its concerns can be anticipated beforehand. This paper has highlighted some of the concerns which might persist for efficient logging in IOT and a possible solution has been presented.

Moreover, how those logs can be stored securely and how the memory can be partitioned to incorporate those logs has been highlighted in this paper.

**References**
[1]  Cong Wang, Qian Wang, Kui Ren, Wenjing Lou, "Privacy-Preserving Public Auditing for Data Storage Security in Cloud Computing", INFOCOM 2010 Proceedings IEEE, Pages 1-9, 14-19 March 2010
[2] John F. Heanue, Matthew C. Bashaw, Lambertus Hesselink, "Volume Holographic Storage and Retrieval of Digital Data", Science Magazine, Vol. 265 no. 5173 pp. 749-752, http://www.sciencemag.org/content/265/5173/749.short, August 1994
[3]  Van der Aalst W, Weijters T, Maruster L, "Workflow mining: discovering process models from event logs", IEEE Transactions on Knowledge and Data Engineering (TKDE), Pages 1128 - 1142, Volume 16, Issue 9, IEEE, Sept. 2004
[4] Animesh K Trivedi, Rajan Arora, Rishi Kapoor, Sudip Sanyal and Sugata Sanyal, "A Semi-distributed Reputation-based Intrusion Detection System for Mobile Ad hoc Networks", arXiv preprint arXiv: 1006.1956, 2010/6/10
[5] Manoj Rameshchandra Thakur, Sugata Sanyal, "A Multi-Dimensional approach towards Intrusion Detection System", arXiv: 1205.2340v1 [cs.CR], 2012/5/10
[6] Ajith Abraham, Ravi Jain, Sugata Sanyal, Sang Yong Han, "SCIDS: a soft computing intrusion detection system", Pages 252-257, Book Distributed Computing-IWDC 2004, 2005/1/1
[7] Animesh Kr. Trivedi, Rishi Kapoor, Rajan Arora, Sudip Sanyal, Sugata Sanyal, "RISM - Reputation Based Intrusion Detection System for Mobile Ad hoc Networks", Third International Conference on Computers and Devices for Communications CODEC-06, pages 234-237, 2006
[8] Feng Xia, Laurence T.Yang, Lizhe Wang and Alexey Vinel, "Internet of Things", INTERNATIONAL JOURNAL OF COMMUNICATION SYSTEMS, Volume 25, Issue 9, pages 1101-1102, September 2012
[9]  Kevin Ashton, "That 'Internet of Things' Thing", RFID journal, June 2009, http://www.rfidjournal.com/articles/view?4986
[10] Kortuem G, Kawsar F Fitton D, Sundramoorthy V, "Smart objects as building blocks for internet of things", Internet Computing IEEE, Volume 14, Issue 1, Pages 44-51, Feb 2010
[11] Garces-Erice L, Felber PA, Biersack EW, Urvoy-Keller G, Ross KW, "Data indexing in peer-to-peer DHT networks", IEEE, Pages 200-208, DOI 10.1109/ICDCS.2004.1281584, 2004
[12] L. Wang, Gregor Laszewski, Marcel Kunze, Jie Tao, "Cloud Computing: A Perspective Study", New Generation Computing- Advances of Distributed Information Processing, pp. 137-146, vol. 28, no. 2, 2008
[13] Sugata Sanyal, Parthasarathi P. Iyer, "Cloud Computing -- An Approach with Modern Cryptography", arXiv preprint arXiv: 1303.1048, 2013/3
[14] Siani Pearson, Yun Shen, Miranda Mowbray, "A Privacy Manager for Cloud Computing", First International Conference, CloudCom, Beijing, China, pp. 90-106, December 2009
[15] Sugata Sanyal, Parthasarathi P. Iyer, "Inter - Cloud Data Security Strategies", arXiv preprint arXiv: 1303.1417, 2013/3
[16] Zoltán Gyöngyi, Hector Garcia-Molina, Jan Pedersen, "Combating web spam with trustrank", VLDB '04 Proceedings of the Thirtieth international conference on Very large data bases - Volume 30 , Pages 576-587, 2004

[17] Christodorescu M, Jha, S, Seshia SA, Song D, Bryant RE, "Semantics-aware malware detection", Security and Privacy, IEEE Symposium, Pages 32 - 46, IEEE, May 2005

[18] Manoj Rameshchandra Thakur, Divye Raj Khilnani, Kushagra Gupta, Sandeep Jain, Vineet Agarwal, Suneeta Sane, Sugata Sanyal, Prabhakar S. Dhekne, "Detection and Prevention of Botnets and malware in an enterprise network", International Journal of Mobile and Wireless Computing, Inderscience, http://arxiv.org/pdf/1312.1629, Volume 5, Issue 2, 2012

[19] Tuhin Borgohain, Amardeep Borgohain, Uday Kumar, Sugata Sanyal, "Authentication Systems in Internet of Things", arXiv preprint arXiv:1502.00870, 2015/2/3

[20] Sugata Sanyal, Ayu Tiwari, Sudip Sanyal, "A multifactor secure authentication system for wireless payment", Book: Emergent Web Intelligence: Advanced Information Retrieval, Pages 341-369, Springer London, 2010/1/1

[21] Wool A, "A quantitative study of firewall configuration errors", Volume 37, Issue 6, Pages 62-67, IEEE Computer Society, IEEE, June 2004

[22] Sotiris Ioannidis, Angelos D. Keromytis, Steve M. Bellovin, Jonathan M. Smith, "Implementing a distributed firewall", Proceedings of the 7th ACM conference on Computer and communications security, Pages 190-199, ACM, 2000

[23] Frederic P. Miller, Agnes F. Vandome, John McBrewster, "Advanced Encryption Standard", Book: Advanced Encryption Standard, ISBN: 6130268297 9786130268299, 2009

[24] Tuhin Borgohain, Uday Kumar, Sugata Sanyal, "Survey of Operating Systems for the IoT Environment", arXiv preprint arXiv:1504.02517, 2015/4/13

[25] Roger M. Needham, Michael D. Schroeder, "Using encryption for authentication in large networks of computers", Communications of the ACM, Volume 21 Issue 12, Pages 993-999, Dec. 1978

[26] Marcus J Ranum, Kent Landfield, Mike Stolarchuk, Mark Sienkiewicz, Andrew Lambeth, Eric Wall, "Implementing a generalized tool for network monitoring", Information Security Technical Report, Elsevier, 1998

[27] Chuck Cranor, Yuan Gao, Theodore Johnson, Vlaidslav Shkapenyuk, Oliver Spatscheck, "Gigascope: high performance network monitoring with an SQL interface", pages 623, DOI 10.1145/564691.564777, ACM, 2002-06-03

[28] J.D. Case, M. Fedor, M.L. Schoffstall, J. Davin, "Simple Network Management Protocol (SNMP)", http://dx.doi.org/10.17487/RFC1157, May 1990

[29] Ferdinand Engel, Kendall S. Jones, Kary Robertson, David M. Thompson, Gerard White, "Network Monitoring", US Patent US6115393 A, Sep 5, 2000